\documentclass[notitlepage,letterpaper, 12pt]{article}
\usepackage[ansinew]{inputenc} 
\usepackage[colorlinks=true,urlcolor=blue,linkcolor=blue,citecolor=blue]{hyperref} 
\usepackage{amsmath}
\usepackage{amsfonts}
\usepackage{amssymb}
\usepackage{graphicx}
\usepackage{subfig}
\usepackage{cancel}
\usepackage{booktabs}
\usepackage{color}
\usepackage{bigints}
\usepackage[top=1cm, bottom=2cm,outer=2cm, inner=2cm]{geometry}
\usepackage{amsthm}
\usepackage{lineno}

\begin{document}
\title {Acceptability Conditions and \\ Relativistic Barotropic \\ Equations of State}
\author{
\textbf{H\'ector Hern\'andez}\thanks{\texttt{hector@ula.ve}}\\
\textit{Escuela de F\'isica, Universidad Industrial de Santander,}\\ 
\textit{Bucaramanga 680002, Colombia} and \\
\textit{Departamento de F\'{\i}sica,}
\textit{Universidad de Los Andes, M\'{e}rida 5101, Venezuela.} \\ 
 \textbf{Daniel Su\'arez-Urango}\thanks{\texttt{danielfsu@hotmail.com}} \\
\textit{Escuela de F\'isica, Universidad Industrial de Santander,  }\\ 
\textit{Bucaramanga 680002, Colombia};\\
and \textbf{Luis A. N\'{u}\~{n}ez}\thanks{\texttt{lnunez@uis.edu.co}} \\
\textit{Escuela de F\'isica, Universidad Industrial de Santander,}\\ 
\textit{Bucaramanga 680002, Colombia} and \\
\textit{Departamento de F\'{\i}sica,}
\textit{Universidad de Los Andes, M\'{e}rida 5101, Venezuela.} \\
}

\maketitle

\begin{abstract}
We sketch an algorithm to generate exact anisotropic solutions starting from a barotropic EoS and setting an ansatz on the metric functions.  To illustrate the method, we use a generalization of the polytropic equation of state consisting of a combination of a polytrope plus a linear term. Based on this generalization, we develop two models which are not deprived of physical meaning as well as fulfilling  the stringent criteria of physical acceptability conditions.

We also show that some relativistic anisotropic polytropic models may have singular tangential sound velocity for polytropic indexes greater than one. This happens in anisotropic matter configurations when the polytropic equation of state is implemented together with an ansatz on the metric functions. The generalized polytropic equation of state is free from this pathology in the tangential sound velocity.

\end{abstract} 

\section{Introduction}
\label{Introduction}
The properties of matter at ultra-high densities $\left(\geq 10^{15} \mbox{gr}/\mbox{cm}^3 \right)$, in the interior of compact objects, have been a subject of study for decades in nuclear physics and astrophysics. Many equations of state (EoS) have been considered in the literature, based on different theoretical models and astronomical observations (see \cite{Glendenning1985, Heiselberg2000, Lattimer2001, Glendenning2000, Hebeler2013, LattimerPrakash2007, LattimerPrakash2016} and references therein). 

The equation of state for supranuclear matter in compact stars plays a fundamental role in the mechanisms triggering supernova explosions and limiting the mass in the formation of black holes and neutron stars. The formulation of an EoS, from a microscopic and experimental point of view, is often postulated from the calculations of two-body potentials and from the nucleon-nucleon dispersion data at densities that go beyond the nuclear saturation density \cite{Akmal1998, Morales2002, Ozel2010}. Regardless of the theory for determining an EoS model, it will always result in a connection between internal pressure and mass-energy density. In the case of non-rotating stars and a certain type of EoS, the structure and hydrostatic equilibrium of these objects will be determined by the constitutive equations, that is, the Tolman-Oppenheimer-Volkoff (TOV) equation and the equation that determines the mass profile within the matter configuration.

In Physics, barotropic EoS relate pressure and density straightforwardly and elegantly. A fluid which is not barotropic is \textit{baroclinic}, i.e., pressure is not the only function of density, and the most paradigmatic example of a baroclinic EoS is the ideal gas equation: $P=NRT/V$. There are other examples of baroclinic fluids: the non-local EoS, 
\[
P(r)=\rho(r)-\frac{2}{r^{3}}\int_{0}^{r}\bar{r}^{2}\rho(\bar
{r})\ \mathrm{d}\bar{r} \,.
\]
 In this type of \textit{non-local} (or \textit{quasi-local}) equation of state a collective behaviour on the physical variables $\rho(r)$ and $P(r)$ is present. The radial pressure $P(r)$ is not only a function of the energy density $\rho(r)$ at that point, but a functional throughout the rest of the configuration. Any change in the radial pressure takes into account the effects of the variations of the energy density within the entire volume (see \cite{HernandezNunezPercoco1999, HernandezNunez2013, HernandezNunez2004, HorvatIIijicMarunovic2011B}, and references therein).

The polytropic equation of state, $P = \kappa \hat{\rho}^{1 + \frac{1}{n}}$, is one of the most venerable barotropic EoS used in the context of Newtonian and relativistic theory, to deal straightforwardly and elegantly with a variety of astrophysical scenarios (see \cite{Glendenning2000, Chandrasekhar1967, Tooper1964, ShapiroTeukolsky1983, NilssonUggla2000} and references therein). 
The pioneering works of Chandrasekhar, Tooper and Kovetz \cite{Chandrasekhar1967, Tooper1964, Kovetz1967} opened the way for the study of relativistic polytropes solving the constitutive equations numerically and determining the physical variables. As discussed in reference \cite{HerreraBarreto2013}, in General Relativity, there are two possible relativistic polytropic equations of state leading to the same Newtonian limit. The difference between them is due to the role played by the baryonic mass density $\hat{\rho}$, or by the total energy density $\rho$. 

Most of the compact object models are considered spherically symmetrical, but the assumption of pressure isotropy can prove to be a somewhat simplifying premise to the matter description. The existence of an anisotropy factor in the local pressures, that is, the possibility that there are two distinct components for the pressure, one radial and the other tangential, induces a more realistic description of the internal structure of a star. On the subject of pressure anisotropy in compact object configuration the literature is numerous: see for example \cite{Ruderman1972, BowersLiang1974, HerreraSantos1997, CosenzaEtal1981, HerreraBarreto2004, HerreraEtal2014, Setiawan2019} and references therein. 

It is worth mentioning that there are several heuristic strategies in introducing anisotropy in relativistic fluids: the earliest one \cite{BowersLiang1974}; quasilocally \cite{DonevaYazadjiev2012};  covariantly \cite{RaposoEtal2019}; the most recent double polytrope \cite{AbellanEtal2020} and finally, providing a metric function or a density profile \cite{Stewart1982, FinchSkea1989, HernandezNunez2004,  HerreraOspinoDiPrisco2008, HernandezNunez2013}. All these strategies have their advantages and disadvantages but have proven to lead to viable models \cite{Setiawan2019}. This physical viability is important because stability and physically acceptability are essential when considering astrophysical scenarios involving self-gravitating matter configurations. In addition to solving the structure equations for a particular set of equations of state, the emerging physical variables have to comply with the several acceptability conditions stated in \cite{DelgatyLake1998, Ivanov2017}. 

In this work, we consider the later of the approaches mentioned above to introduce anisotropy in relativistic bounded matter configuration, i.e. to provide a barotropic EoS and an ansatz on the energy density profile. We assumed a polytrope for the barotropic EoS and found that, within this approach, anisotropic polytropes could have a singular tangential sound velocity at the surface of the matter distribution when the polytropic index is $n > 1$. Thus, we implemented a generalization of the polytropic equation of state which consists of a combination of a polytrope plus a linear and an independent density terms. Along this line, we developed an algorithm to generate exact anisotropic polytropic solutions, beginning with an ansatz on the density profiles. Thus, with this scheme we avoided cumbersome and counter-intuitive change of variables. 

This paper is organized as follows. Section \ref{FieldEquations} describes the general equation framework of General Relativity. In Section \ref{PhysicalAcceptabilityConditions} we list the set of acceptable conditions adhered by our models; next Section \ref{AllEquationState} describes the algorithm to obtain any polytrope, starting from an energy density profile. In Section \ref{BarotropicGeneralities} we implement the generalization of a polytropic EoS and show that both relativistic versions have the same Newtonian limit. In Section \ref{GeneratingMetric}, we model two matter distributions starting from two physical interesting density profiles. Next, in Section \ref{AnalyticalAcceptability}, we discuss the acceptability conditions and finally  in Section \ref{FinalRemarks} we wrap-up our final remarks.

\section{The field equations}
\label{FieldEquations}
Let us consider the interior of a dense star described by  a spherically symmetric  space-time line element written as
\begin{equation}
\mathrm{d}s^2 = {\rm e}^{2\nu(r)}\,\mathrm{d}t^2-{\rm e}^{2\lambda(r)}\,\mathrm{d}r^2- r^2 \left(\mathrm{d}\theta^2+\sin^2(\theta)\mathrm{d}\phi^2\right),
\label{metricSpherical}
\end{equation}
with regularity conditions at 
$r=r_c=0$, i.e. ${\rm e}^{2\nu_c}=$ constant,  ${\rm e}^{-2\lambda_c}= 1$, and $\nu^{\prime}_c=\lambda^{\prime}_c=0$. 

Additionally, the interior metric should match continuously the Schwarzschild exterior solution at the surface of the sphere,  $r=r_b$. This implies that ${\rm e}^{2\nu_b}={\rm e}^{-2\lambda_b}=1-\mu = 1 -2M/r_b$,  where $M$ is the total mass and $\mu=2M/r_b$ the compactness of the configuration.  From now on, subscripts $b$ and $c$ indicate the evaluation of a particular variable at the boundary and at the center of the matter distribution, respectively.  

We shall consider a distribution of matter consisting of a non-Pascalian fluid represented by an energy-momentum tensor:
\begin{equation}
T_\mu^\nu = \mbox{diag}\left[\rho(r),-P(r),-P_\perp(r),-P_\perp(r)  \right] \,,
\label{tmunu}
\end{equation}
where $\rho(r)$ is energy density, with $P(r)$ and $P_\perp(r)$ the radial and  tangential pressure, respectively. 

From the Einstein field equations we obtain these physical variables in terms of the metric functions as
\begin{eqnarray}
\rho(r)&=& \frac{ {\rm e}^{-2\lambda}\left(2 r \lambda^{\prime}-1\right)+1 }{8\pi r^{2}}\,,\label{FErho} \\
P(r) &=&  \frac {{{\rm e}^{-2\,\lambda}}\left(2r \,\nu^{\prime} +1\right) -1}{8 \pi\,{r}^{2}}\,\label{FEPrad} \qquad \textrm{and} \\
P_\perp(r) &=&-\frac{{\rm e}^{-2\lambda}}{8\pi}\left[ \frac{\lambda^{\prime}-\nu^{\prime}}r-\nu^{\prime \prime }+\nu^{\prime}\lambda^{\prime}-\left(\nu^{\prime}\right)^2\right] \label{FEPtan} \,, 
\end{eqnarray}
where primes denote differentiation with respect to $r$. 

Now assuming 
\begin{equation}
    m(r)=\frac{r}{2} \left(1- {\rm e}^{-2\lambda} \right)\,,
\end{equation}
the Tolman-Oppenheimer-Volkoff equation -- i.e. $T^{\mu}_{r \; ; \mu}~=~0$, the hydrostatic equilibrium equation-- for this anisotropic fluid can be written as   
\begin{equation}
\frac{\mathrm{d} P}{\mathrm{d} r} +(\rho +P)\frac{m + 4 \pi r^{3}P}{r(r-2m)} -\frac{2}{r}\left(P_\perp-P \right) =0\,,
 \label{TOVStructure1}
\end{equation}
and together with
\begin{equation}
\label{MassStructure2}
\frac{\mathrm{d} m}{\mathrm{d} r}=4\pi r^2 \rho \,,
\end{equation}
constitute the relativistic stellar structure equations. 

Clearly, it is equivalent to solve the Einstein system (\ref{FErho})-(\ref{FEPtan}) or to integrate the structure equations (\ref{TOVStructure1})-(\ref{MassStructure2}). In the first case we obtain the physical variables $\rho(r)$, $P(r)$ and $P_\perp(r)$ provided the metric functions $\lambda(r)$ and $\nu(r)$, while in the second approach  we integrate the structure equations (\ref{TOVStructure1})-(\ref{MassStructure2}) giving two barotropic equations of state, $P=P(\rho)$ and $P_{\perp}=P_{\perp}(P(\rho),\rho) \equiv P_{\perp}(\rho)$. 

These two EoS involving the radial and tangential pressures, together with the matching conditions --initial conditions for the system of first-order differential equations--, $P(r_b)=P_{b}=0$ and $m(r_b)=m_{b}=M$, lead to a system of differential equations for $\rho(r)$ which can be solved to obtain the inner structure of a self-gravitating relativistic compact object. 

As is well known, stellar compact objects have been modelled for decades as Pascalian fluids, that is, with an isotropic pressure distribution. However, a considerable number of studies have shown that the pressures within compact objects could be anisotropic, i.e. non-Pascalian fluids with unequal radial and tangential pressures,  $\Delta\equiv P_\perp-P\neq 0$ \cite{BowersLiang1974, HerreraSantos1997, Setiawan2019, RaposoetEtal2019}, and it can influence the stability of the compact object --inducing cracking or overturning--, its mass-radius ratio, or/and its maximum mass (see \cite{Herrera1992, DiPriscoHerreraVarela1997, AbreuHernandezNunez2007b, GonzalezNavarroNunez2017} and references therein, particularly, reference \cite{HerreraSantos1997}).

\section{Physical acceptability conditions}
\label{PhysicalAcceptabilityConditions}
Stability is a crucial concept when considering self-gravitating stellar models: only objects in stable equilibrium are of astrophysical interest.
Thus, in addition to solving the structure equations (\ref{TOVStructure1}) and (\ref{MassStructure2}) for a particular set of equations of state (i.e. $P=P(\rho)$ and $P_{\perp}=P_{\perp}(\rho)$), the emerging physical variables have to comply with the several acceptability conditions \cite{DelgatyLake1998}. As B.V. Ivanov \cite{ Ivanov2017} recently showed, there are several independent acceptability conditions to be fulfilled by any general relativistic anisotropic model of a compact object.

In this work, ``physically acepted'' models are those which comply with the following nine conditions:
\begin{enumerate}
\item[{\bf C1}] $2m/r < 1$; this implies
    \begin{enumerate}
    \item that the metric potentials $\textrm{e}^{\lambda}$ and $\textrm{e}^{\nu}$ are positive, finite and free from singularities within the matter distribution, satisfying $\textrm{e}^{\lambda_{c}} = 1$ and $\textrm{e}^{\nu_{c}}=$ constant at the center of the configuration;
    \item the inner metric functions match to the exterior Schwarzschild solution at the boundary surface;
    \item the interior redshift should decrease with the increase of $r$ \cite{Buchdahl1959,Ivanov2002B};
    \end{enumerate}

\item[{\bf C2}] Positive density and pressures, finite at the center of the configuration with $P_c=P_{\perp c}$ \cite{Ivanov2002B};
\item[{\bf C3}] $\rho^{\prime} < 0$, $P^{\prime} < 0$ $P_{\perp}^{\prime} < 0$ with density and pressures having maximums at the center, thus $\rho^{\prime}_{c}=P^{\prime}_{c} = P^{\prime}_{\perp c}=0$,  with $P_{\perp} \geq P$;
\item[{\bf C4}] The strong energy condition for imperfect fluids: $\rho - P - 2P_{\perp} \geq 0$  \cite{KolassisSantosTsoubelis1998,PimentelLoraGonzalez2017};
\item[{\bf C5}]  The dynamic perturbation analysis restricts the adiabatic index \cite{HerreraSantos1997,HeintzmannHillebrandt1975,ChanHerreraSantos1993,ChanHerreraSantos1994} 
\[
\Gamma = \frac{\rho + P}{P} v_s^{2} \geq \frac{4}{3} \,.
\]
\item[{\bf C6}] Causality conditions on sound speeds: $0 < v_{s}^2 \leq 1$ and $0 < v_{s \perp}^2 \leq 1$;
\item[{\bf C7}] The Harrison-Zeldovich-Novikov stability condition: $\mathrm{d}M(\rho_c)/\mathrm{d}\rho_c > 0$ \cite{HarrisonThorneWakano1965,ZeldovichNovikov1971};

\item[{\bf C8}] Cracking instability against local density perturbations, $\delta \rho = \delta \rho(r)$, briefly described in the next section and in references \cite{GonzalezNavarroNunez2017,GonzalezNavarroNunez2015,HernandezNunezVasquez2018};
\item[{\bf C9}] The adiabatic convective stability condition  $\rho^{\prime \prime} \leq 0$, which is more restrictive than the outward decreasing density and pressure profiles \cite{HernandezNunezVasquez2018}.
\end{enumerate}

Notice that the standard $2m/r < 1$ condition is different from the stronger $(m/r)^{\prime} > 0$, as required by B.V. Ivanov in \cite{Ivanov2017}. Clearly, if $(m/r)^{\prime} > 0$ we obtain well behaved metric functions but, there are cases with $(m/r)^{\prime} < 0$ having physically reasonable metric coefficients \cite{SuarezurangoHernandezNunez2020}.  Thus, $(m/r)^{\prime} > 0$ should be considered as a sufficient but not a necessary condition to obtain ``well behaved'' metric potentials.

The restriction $0 \geq P_{\perp}^{\prime} \geq P^{\prime}$ also presented in \cite{Ivanov2017}, implies the most simple cracking condition, i.e. $-1 \leq v^{2}_{s_\perp} -v_s^2 \leq 0$ \cite{Herrera1992,AbreuHernandezNunez2007b}. But our models include a more elaborate cracking criterion with variable local density perturbations, $\delta \rho~=~\delta \rho(r)$ described in references \cite{GonzalezNavarroNunez2017,GonzalezNavarroNunez2015} and \cite{HernandezNunezVasquez2018} (see Appendix A-2 for a discussion).

Ivanov, also, correctly requires that $P_{\perp} \geq P$ and we shall show that, at least for the polytropic EoS, this condition leads to more stable matter configurations.

In the next sections we shall discuss the impact of barotropic and baroclinic fluids to model anisotropic compact objects.

\section{All Barotropic Equations of State}
\label{AllEquationState}
In general, barotropic fluids are those where the pressure is the only function of  density, i.e., $P=P(\rho)$, and vice-versa. Although they may be considered unrealistic, their simplicity motivates a pedagogical value in illustrating the several approaches used to solve different systems and  ``physically'' interesting scenarios.

In this work we shall consider the ``pedagogical'' barotropic EoS for radial pressure, $P~=~P(\rho)$ and from this assumption formally integrate $\nu^{\prime}$ from equation (\ref{FEPrad}) as,
\begin{equation}
\nu(r) = \frac12 \int \frac{1}{r} 
\left\{\textrm{e}^{2\lambda}\left[8\pi \title{r}^2P(\rho) +1 \right] -1  \right\} \textrm{d}r + \mathcal{\mathbf{C}} \,,   
\label{Integnu}    
\end{equation}
where $\mathcal{\mathbf{C}}$ is an integration constant that can be obtained from the condition: ${\nu(r_b)}={-\lambda(r_b)}$. A similar equation is reported in \cite{AbellanEtal2020},  but in our case we shall implement this formal integration by considering any barotropic  EoS, $P=P(\rho)$, as an input. Given any barotropic EoS and a $\lambda$ function --so it is possible to integrate equation (\ref{Integnu})--, we can determine the baroclinic EoS for the tangential pressure via equation (\ref{FEPtan}), (or equivalently, from the TOV equation (\ref{TOVStructure1})) as
\begin{equation}
P_\perp= P(\rho) +\frac{{{\rm e}^{2\lambda}} }{4}{ { \left( \rho+P(\rho)\right) \left( 8\,\pi\,{r}^{2}{ P(\rho)} -{{\rm e}^{-2\lambda}} +1 \right) }} +\frac{r}{2} v_s^2 \rho^{\prime} \,,
\label{Ptbaro}
\end{equation}
where $v_s^2$ is the  radial sound velocity. The  tangential sound velocity can be expressed through a more complex relation, as
\begin{equation}
v_{s_\perp}^2 = \frac{{\rm d}P_\perp}{\rm{d} \rho} = \frac12\left[ 3  +r\frac{\rho^{\prime \prime}}{\rho^{\prime}} \right] v_s^2 + \frac{1}{\rho^{\prime}}\left[\frac{{{\rm e}^{2\lambda}} }{4}{ { \left( \rho+P(\rho)\right) \left( 8\,\pi\,{r}^{2}{ P(\rho)} -{{\rm e}^{-2\lambda}}+1 \right) }} \right]^{\prime} + \frac{r}{2} (v_s^2)^{\prime}
\,.
\label{veltbar}
\end{equation}

The input barotropic EoS can be either an analytic relation between the pressure and the density, $P \leftrightarrow \rho$, or a more realistic   ``numeric'' relation. In the next section we work out several cases with analytic EoS, implementing an exact integration of equation (\ref{Integnu}) (with $P(\rho,r)$ depending on $r$, $\lambda$ and $\lambda^{\prime}$). We carry out the analytic barotropic example by using a generalization of the polytropic EoS of the form $P = \kappa {\rho}^\gamma+\alpha \rho -\beta $ with $\kappa$, $\alpha$ and $\beta$  parameters to be determined. This generalization avoids some pathologies that, we will prove exist in all 
``standard'' polytropic non-Pascalian fluids. Those models having  $\alpha~=~\beta~=~0$, with $1~<~\gamma~<~2$, unavoidably present a singularity in the tangential sound velocity (\ref{veltbar}) at the boundary of the matter distribution. 

\section{A generalized polytropic equation of state}
\label{BarotropicGeneralities}
The polytropic EoS deals with a variety of physically attractive astrophysical scenarios \cite{HerreraBarreto2013}. As we have mentioned, we ``need'' a barotropic EoS and a particular density profile, i.e. a specific $\lambda(r)$. 

In this section, we shall discuss the barotropic polytropic EoS as one of the examples considered to integrate the equation (\ref{Integnu}). We consider an equation that includes polytropes and what we will call the ``master'' equation of state: 
\begin{equation}
    P = \kappa \hat{\rho}^{\gamma}+\alpha\hat{\rho}-\beta \,,
    \label{polinewt1}
\end{equation}
where $P$, $\hat{\rho} =\mathcal{N}m_{0}$ , $\kappa$ and $\gamma=1+1/n$ are: the isotropic pressure, the (baryonic) mass  density, the polytropic constant and $n$ the  polytropic index, respectively. Notice that the number of particles is $\mathcal{N}$, while $m_{0}$ represents the baryonic particle mass.  
Observe that $\kappa, \alpha$ and $\beta$ are non-independent parameters related at the boundary surface as:
\begin{equation}
  \beta=\kappa \hat{\rho_b}^{\gamma}+\alpha \hat{\rho}_b \,.
  \label{betamaster}
\end{equation}

In section \ref{GeneratingMetric} we use two distinct ``seeds'' $\lambda(r)$-function: a Tolman VII-like seed-metric \cite{DelgatyLake1998,Tolman1939} and  a generalization of Buchdahl's  one-parameter  solution \cite{DelgatyLake1998,Buchdahl1959}. In this case $\nu$  -- equation (\ref{Integnu})-- can be obtained analytically.

\subsection{One Newtonian and two relativistic polytropes}
Following \cite{HerreraBarreto2013}, we briefly present both cases for our ``master'' polytropic EoS (\ref{polinewt1}):
\begin{enumerate}
    \item In the first case we are considering the baryonic particle density, $\mathcal{N} =\hat{\rho} / m_0$ with $m_0$ the baryonic mass.  The equation of state (\ref{polinewt1}) is combined with the adiabatic first law of thermodynamics, resulting in:
  \begin{equation}
\mathrm{d}\left(\frac{\rho}{\mathcal{N}} \right) + P\mathrm{d}\left(\frac{1}{\mathcal{N}} \right) = 0 \,\, \Rightarrow \,\,
\frac{\mathrm{d}}{\mathrm{d}{\hat \rho}} \left(\frac{\rho}{\hat \rho}\right) =
\frac{P}{ {\hat \rho}^{2}} \quad \Rightarrow 
\frac{1}{{\hat \rho}}\frac{\mathrm{d} \rho}{\mathrm{d}{\hat \rho}}-\frac{\rho}{{\hat \rho}^2}= \frac{P}{ {\hat \rho}^{2}} \,,
    \label{firstlaw1}
\end{equation}
thus
\begin{equation}
\frac{\mathrm{d} \rho}{\mathrm{d}{\hat \rho}}-\frac{\rho}{{\hat \rho}} =    \kappa{\hat \rho}^{\gamma-1}+{\alpha}-\frac{\beta}{{\hat \rho}}   \,,
  \label{firstlaw1b}
\end{equation}
where $\gamma=1 + \frac{1}{n}$ is the polytropic exponent. 
Then equation  (\ref{firstlaw1b}) can be integrated and, by using  (\ref{polinewt1}), we obtain two possible solutions 
\begin{equation}
\left\{
\begin{array}{lllll}
\gamma \neq 1   &  \,\, \Rightarrow \,\, &\rho = \dfrac{\kappa {\hat \rho}^\gamma }{\gamma-1}+\left[\alpha \ln({\hat \rho})+ C_{1} \right]{\hat \rho} +\beta & \\
                &    &  &   \\
\gamma = 1  &  \,\, \Rightarrow \,\,  & \rho = \left[(\alpha+\kappa) \ln({\hat \rho})+C_{1}\right]{\hat \rho} +\beta&
\end{array}
\right.
\label{solcase11}
\end{equation}
where $C_{1}$ is a constant of integration.

\item The second approach takes into account the energy density $\rho$ and starts with 
\begin{equation}
    P = \kappa \rho^{\gamma}+\alpha \rho -\beta \,,
    \label{politroRho}
\end{equation}
so that equation (\ref{firstlaw1}) leads to
\begin{equation}
\frac{\mathrm{d} \rho}{\mathrm{d}{\hat \rho}}-\frac{1}{{\hat \rho}}\left[(1+\alpha)\rho-\beta\right] = \frac{\kappa}{{\hat \rho}}\ {\rho}^{\gamma} \,,
\end{equation}
and formally we get
\begin{equation}
    \int \frac{\mathrm{d} \rho}{\kappa{ \rho}^{\gamma}+(1+\alpha)\rho-{\beta}}=\ln{\left[\frac{{\hat \rho}}{C}\right]} \qquad \textrm{with} \quad \gamma \neq 1 \,.
\end{equation}
Note that when $\alpha=\beta=0$ we obtain 
\begin{equation}
    \rho= \frac {{\hat \rho}}{{\it C}} \left[ 1-\kappa\left(\frac {{\hat \rho}}{ C}\right)^{\frac1n}
 \right] ^{-n} = \frac{{\hat \rho}}{\left[ C^{\frac1n} -\kappa {\hat \rho}^{\frac1n}\right]^n} \,,
\end{equation}
as in \cite{HerreraBarreto2013}, and if $\gamma= 1$, the result is
\begin{equation}
\rho= C{\hat \rho}^{1+\alpha+ \kappa} + \frac {\beta}{1+\alpha +\kappa} \,,
\end{equation}
where $C$ is the constant of integration.
\end{enumerate}

\subsection{The relativistic ``master'' polytropic equation of state}
Following the second approach for relativistic polytropes --using the energy density $\rho$ and not the baryonic mass density $\hat{\rho}$--, we shall generalize (\ref{politroRho}) in the form of
\begin{equation}
    P =\kappa {\rho}^{1 + \frac{1}{n}}+\alpha \rho -\beta 
    = \kappa \left[\frac{ e^{-2\lambda}\left(2 r \lambda^{\prime}-1\right)+1 }{8\pi r^{2}} \right]^{1 +\frac{1}{n}} + \alpha\left[{\frac{ e^{-2\lambda}\left(2 r \lambda^{\prime}-1\right)+1 }{8\pi r^{2}}}\right] -\beta \,.
\label{MasterPolytropic}
\end{equation}

From equation (\ref{MasterPolytropic}), and the fact that on the surface the radial pressure vanishes, we have
\begin{equation}
  \beta=\kappa {\rho_b}^{1 + \frac{1}{n}}+\alpha \rho_b  \,,
  \label{betamasters}
\end{equation}
 with
\begin{equation}
\kappa=  
\frac{\sigma - \alpha\left[1 - \varkappa \right]}{ {\rho_c}^{\frac{1}{n}}\left[1 - \varkappa^{1 + \frac{1}{n}}\right] }
\,,
\label{kmaster}
\end{equation}
where
\begin{equation}
    \sigma= \frac{P_c}{\rho_c} \quad \mbox{and} \quad \varkappa =\frac{ \rho_b}{\rho_c} \,.
\label{sigma}
\end{equation}

By using equation (\ref{MasterPolytropic}) in the expression of the radial and the tangential sound velocities we obtain
\begin{eqnarray}
v_s^2 &=&\kappa \left[ 1+\frac{1}{n} \right] \rho^{\frac{1}{n}} +\alpha  \quad \textrm{and} \label{velradgen} \\
v_{s_\perp}^2 &=& \frac12\left[3+r\frac{\rho^{\prime \prime}}{\rho^{\prime}}  \right] v_s^2+\frac{1}{\rho^{\prime}}\left[\frac{{{\rm e}^{2\lambda}} }{4}{ { \left( \rho+P\right) \left( 8\,\pi\,{r}^{2}{ P} -{{\rm e}^{-2\lambda}}+1 \right) }} \right]^{\prime} 
+ \frac{\kappa r(n+1)\rho^{\frac{1}{n}}}{2n^2\rho}{\rho^{\prime}}\, ,
\label{veltgen}
\end{eqnarray}
respectively. 

Observe that the parameters in equation (\ref{MasterPolytropic}) are physically meaningful. Some are simple relations, like $\sigma = P_c/\rho_c$,  describing how rigid is the centre of the matter distribution. Others, like $\varkappa =  \rho_b/\rho_c$, sketches the density drop from the centre to the surface of the compact object, and from equation (\ref{velradgen}) it is clear that $\alpha$ is related to the causality of the radial sound velocity $0~<~v_s^2~<~1$. There are more complex relation like $\kappa$ given by equation (\ref{kmaster}) or $\beta$ by equation (\ref{betamasters}). 

At this point, it is evident from expression (\ref{veltgen}) that, if the density vanishes at the surface of the distribution, the tangential sound velocity becomes singular at the boundary of the distribution for any polytropic index $n > 1$. This is a general result when  ``standard'' polytropic EoS are implemented together with the strategy to provide an educated guess on the metric functions and was the rationale for introducing a ``master'' polytropic EoS (\ref{MasterPolytropic}). 

Note that if $n\rightarrow \infty$ the master equation becomes a linear EoS: $P=(\kappa+\alpha)\rho-\beta$, the equation (\ref{kmaster}) results in  $\sigma=\kappa(1-\varkappa)$, the sound velocity tends to $v_s^2=\kappa+\alpha$ and the last term of (\ref{veltgen}) vanishes.

Clearly, we can see then that the  ``master'' barotropic equation (\ref{MasterPolytropic}) includes the following particular cases:
\begin{enumerate}
\item For $\alpha=\beta=0$, we get back the standard polytropic EoS:
\begin{equation}
    P = \kappa {\rho}^{1 + \frac{1}{n}} \,,
    \label{polytropic}
\end{equation}  
where both $\kappa = \sigma/\rho_{c}^{\frac{1}{n}} $ and $n$ are positive parameters. The vanishing $\beta$ implies 
\begin{equation}
\rho(r_b)=P(r_b)=P_\perp(r_b)=0 \,,
\label{rhobpoly0}
\end{equation}
at the boundary of the distribution. Additionally, the radial sound velocity can be written as:
\begin{equation}
v_s^2 =\frac{\kappa (n+1)}{n} \rho^{\frac{1}{n}}\,. 
\end{equation}

As we have pointed out above, observe from equations (\ref{veltgen}) and (\ref{rhobpoly0}), that the tangential sound velocity --for a ``standard'' polytropic non Pascalian fluid-- clearly diverges. At the boundary surface, $r = r_b$, the last term in equation (\ref{veltgen}) becomes infinite for $n > 1$.   The tangential sound velocity defined as (\ref{veltgen}) will be singular at the boundary surface of the distribution and this is frequently overlooked in the literature (see, for example, references \cite{ThirukkaneshRagel2012,Ngubelanga2015, AbellanEtal2020, TakisaMaharaj2013, Malaver2015, NgubelangaMaharaj2017, SharifSadiq2018}). This is the main motivation for introducing a more general polytropic anisotropic equation of state. 

\item When $n = 1$, equation (\ref{MasterPolytropic}) corresponds to a quadratic equation of state, i.e. 
\begin{equation}
   P = \kappa \rho^2 +\alpha\rho -\beta \,,
   \label{quadratic}
\end{equation}    
with, the radial sound velocity
\begin{equation}
v_s^2=2\kappa \rho + \alpha \,, 
\end{equation}    
and the polytropic parameter depending on 
\begin{equation}
\kappa=  \frac{\sigma - \alpha\left[1 - \varkappa \right]}{ {\rho_c} \left[1 - \varkappa^{2}\right] }\,.
\label{kquadra}
\end{equation}
This case was considered by Feroze and Siddiqui \cite{FerozeSiddiqui2011} when, using the Durgapal-Bannerji scheme \cite{Durgapal1982}, they generalized a previous work by Maharaj \cite{ThirukkaneshMaharaj2008, MaharajTakisa2012}. In these two papers the authors show that their anisotropic solutions could describe a possible compact object. 

\item  If  $\beta =0$, this EoS is known as the {\it generalized equation of state} \cite{Chavanis2014}:
\begin{equation}
    P = \kappa {\rho}^{1 + \frac{1}{n}} + \alpha \rho  \,.
    \label{polygen}
\end{equation}
A simple inspection shows that $P = 0$ on the surface implies that
\begin{equation}
{\alpha}=-\kappa{\rho_b}^{\frac1n} \,,
\label{kpolgen}
\end{equation}
and the radial sound velocity can be written as
\begin{equation}
v_s^2 =\kappa\left[\frac{ (n+1)}{n} \rho^{\frac{1}{n}} -{\rho_b}^{\frac1n}\right]\,. 
\end{equation}
Thus, the parameters $\kappa$ and $\sigma$ are related as follows
\begin{equation}
\kappa= \frac{\sigma}{{\rho_c}^{\frac{1}{n}}\left[1-{\varkappa}^{\frac{1}{n}}\right] } \,.
\label{kpolilin}
\end{equation}

Equation (\ref{polygen}) has been used for cosmological purposes, with values $n~>~0$. In the early universe, when the energy density was high, the polytropic part dominates over the linear component \cite{Chavanis2014}. Later, in a series of works \cite{AzamEtal2016,MardanEtal2018, NoureenEtal2019} this EoS was implemented to model charged and neutral anisotropic polytropes. 

\item Obviously, in the case of $\kappa=0$, our master equation (\ref{MasterPolytropic}) corresponds to a linear state equation: 
 \begin{equation}
   P = \alpha\rho -\beta \,, 
   \label{lineal}
\end{equation} 
where $\beta=\alpha \rho_b$, now the radial sound speed is constant: $v_s^2=\alpha$ and we have
\begin{equation}
\alpha= \frac{\sigma }{1 - \varkappa } \,.
\label{klinear}
\end{equation}
Nilsson and Uggla  \cite{NilssonUggla2000b}  worked out a detailed study for an static, perfect fluid spherical distribution of matter described by: $(\eta -1)P=\rho -\rho_0$, where the constants $\rho_0$ and $\eta$  satisfy $\rho_0 \geq 0$ and  $\eta \geq 1$. Later on, P.H.~Chavanis \cite{Chavanis2008}  analyses the structure and stability of compact objects with linear EoS. In reference \cite{LaiXu2009} the authors, considering the framework of the bag model, describe cold quark matter by using a linear EoS of the type of $P=k(\rho-\rho_b)$,  with two free parameters $k$ and $\rho_b$. 

\item Recently, K.N. Singh and collaborators \cite{SinghEtal2020}, based on some dubious motivations, discussed several viable polytropic anisotropic models with $\alpha = 0$.

\end{enumerate}

\section{Analytical anisotropic polytropic solutions}
\label{GeneratingMetric}
As discussed previously, to integrate (\ref{Integnu}) we need to provide both, the equation of state $P(\rho)$ and the ``mass function''  $\textrm{e}^{2\lambda(r)}$ --or the density profile. Also observe that, by using equation (\ref{MasterPolytropic}), we can split the integral (\ref{Integnu}) in two parts
\begin{equation}
 \nu=  \int \left[4\pi\, r \left( \alpha\,\rho  -\beta \right){{\rm e}^{2\,\lambda}} -\frac {1-{{\rm e}^{2\,\lambda }}}{2r}\right] {\rm d}r+ 4\pi \kappa \int r{{\rm e}^{2\,\lambda  }} \rho ^{1+\frac{1}{n}}\,{\rm d}r+ \mathcal{\mathbf{C}}\,,
 \label{nuMaster}
\end{equation}
this last expression can be very useful when considering particular cases of the equation of state, such as those shown in the previous section.

In the present work we have selected two well known $\lambda(r)$-seed-functions generating a pair of reasonable density profiles:
\begin{eqnarray}
\mbox{\bf Seed 1:}& & 
{\rm e}^{2\lambda}= \left[1+Ar^2 + Br^4\right]^{-1}\,\, \Rightarrow \,\,\rho= -\dfrac {3A + 5B{r}^2 }{8\pi}\,  \label{seed1} \\ 
\textrm{and,} & & \nonumber \\ 
\mbox{\bf Seed 2:}& & {\rm e}^{2\lambda}=\dfrac {K(1+A{r}^{2})}{K+B{r}^{2}}\,\, \Rightarrow \,\, \rho=\frac { \left(K A-B \right)  \left(3+A{r}^{2} \right) }{8\pi K 
 \left(1+A{r}^{2}\right) ^{2}}
 \,.
\label{seed2}
\end{eqnarray}
Here $A$, $B$ and $K$ are parameters determined from the matching conditions, $P(r_b)=0$ and $m(r_b)=M$.

The $\lambda(r)$-seed-function 1  corresponds to a two-parameters Tolman VII-like metric element \cite{Tolman1939,DelgatyLake1998}. This family of solutions is one of the most frequent parabolic density profiles considered in stable models of neutron stars (see \cite{RaghoonundunHobill2015,BharMuradPant2015,AzamMardanRehman2015,Raghoonundun2016}). From the $\lambda$-function (\ref{seed1}) we obtain the condition 
\begin{equation}
r\frac{\rho^{\prime\prime}}{\rho^{\prime}}=1 \,
\end{equation}
in equation (\ref{veltgen}), which is consistent with criteria {\bf C2} and {\bf C9}.

The second  $\lambda$-function (\ref{seed2}) is a generalization of Buchdahl's  one-parameter  solution \cite{DelgatyLake1998,Buchdahl1959}, obtained when $B =-A$ and $K=2$. Now, the constraint for the derivative of the density profile turns to be
\begin{equation}
 r\frac{\rho^{\prime\prime}}{\rho^{\prime}}=   {\frac {5 -\left(22+3{A}{r}^{2}\right)A{r}^{2}}{ \left(5+ A{r}^{2} \right) \left(1+ A{r}^{2} \right) }}\, .
\end{equation} 
The two parameters $A$ and $B$ can be determined as
\[
\lambda=\lambda(A,B,r) \,\, \Rightarrow \,\, \rho(A,B,r) \,\, \Rightarrow \,\,
\left\{\begin{array}{l}
P(A,B,r_b)=0 \,\, \Rightarrow \,\, A \\
\quad m(B,r_b)=\mu\, r_b/2 \,\, \Rightarrow \,\, B \, .
\end{array}
\right.
\]
Notice that instead of the total mass $M$ we have used the compactness parameter $\mu =2M/r_b$ and that in the case of the standard polytropic equation (\ref{polytropic}), we have $P(r_b)=\rho(r_b)=P_\perp(r_b)=0$.

Some exact solutions of Einstein's equations may contain only one free parameter in the metric functions which emerges from the boundary conditions, as
\begin{equation*}
\lambda=\lambda(C,r) \,\, \Rightarrow \,\, \rho(C,r) \,\, \Rightarrow \,\, C  \,\, \Leftarrow \,\, P(C,r_b)=0\,, 
\end{equation*}
then $\mu$ should have a unique precise value for any polytropic index $n$.
The requirement to have an exact valued for compactness $\mu$ of any polytropic index $n$ is sometimes overlooked. In reference \cite{ThirukkaneshRagel2012}, the authors use $\mu = 0.96$ instead of the correct $\mu$ considered in \cite{FinchSkea1989}.

In the next sections, we shall use equation (\ref{MasterPolytropic}) --and two $\lambda$-seed-functions, (\ref{seed1}) and (\ref{seed2})-- to model spherically symmetric and anisotropic compact objects. This scheme starts by providing a particular density profile and not a cumbersome and superfluous changes of  variables to solve the field equations analytically.
\begin{figure}[t!]
\centering
\includegraphics[width=3.2in,height=2.4in]{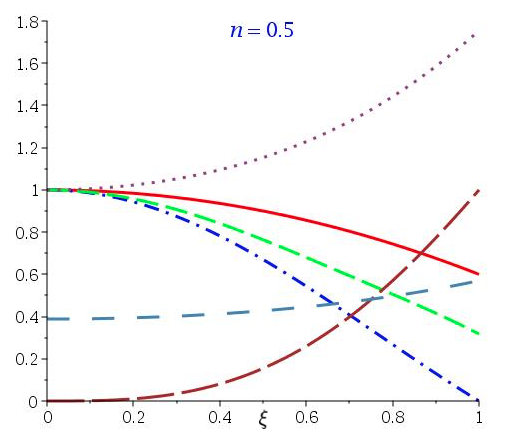}
\includegraphics[width=3.2in,height=2.4in]{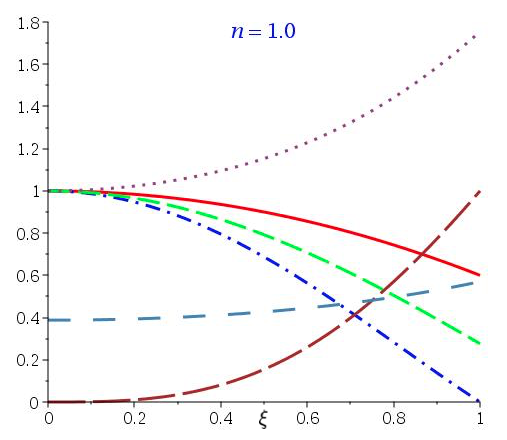}\\
\includegraphics[width=3.2in,height=2.4in]{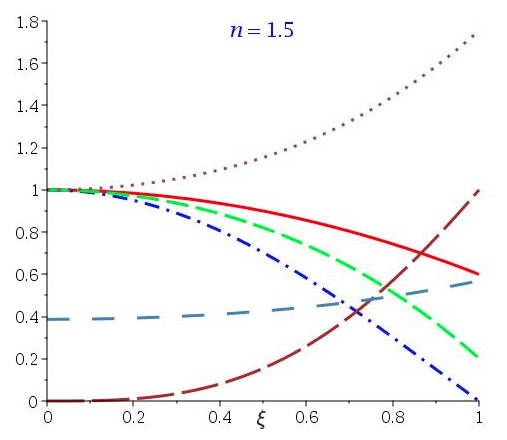}
\includegraphics[width=3.2in,height=2.4in]{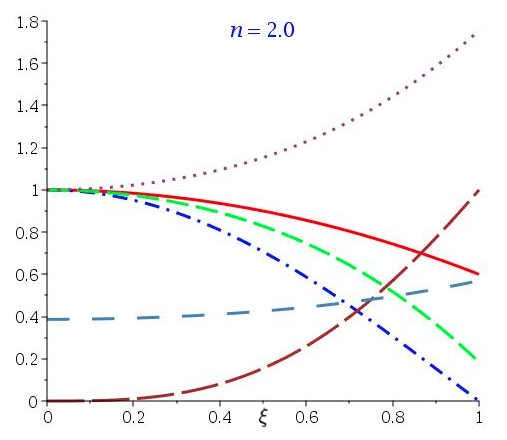}\\
\includegraphics[width=2.6in,height=0.4in]{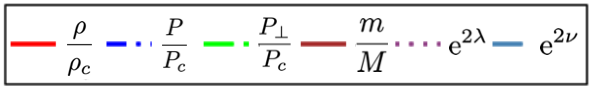}
\caption{Physical variables and metric functions for a matter configuration with EoS-1.  The various plates illustrate different values of $n$ having $\sigma = 0.12$. The normalised distributions of density, radial/tangential pressures, mass and components of the metric tensor are given in terms of the dimensionless radius $\xi=r/r_b$. All the physical and geometrical variables are well behaved and comply with acceptability criteria {\bf C1}, {\bf C2}, and {\bf C3}.} 
\label{EoS1Functions}
\end{figure}

\subsection{Modeling the anisotropic, polytropic EoS-1}
\label{Modeling}

To build the first model, we integrate equation (\ref{nuMaster}) by using equation (\ref{seed1}), and obtain  
\begin{equation}
\nu= \frac{\lambda\left(1+ 5\alpha\right)}{4} + \frac {A(1+\alpha)+16\pi\beta }{4\sqrt {{A}^{2}-4\,B}} {\rm arctanh}\left[{\frac {A+2B {r}^{2}}{\sqrt{{A}^{2}-4B}}}\right] 
 +  4\pi \kappa \int r{{\rm e}^{2\,\lambda  }} \rho ^{1+\frac{1}{n}}\,{\rm d}r +\mathcal{\mathbf{C}} \, .
 \label{nueos1}
\end{equation}
In the Appendix we have sketched the integration strategy for  $n = 0.5$, $1.0$, $1.5$ and $2.0$. 

Next, from the matching conditions $P(r_b)=0$ and $m(r_b)=M$ we solve two of the five unknown parameters:
\[
\beta=k\rho_b^{1+\frac{1}{n}}+\alpha \rho_b \quad \textrm{and} \quad A=-\frac{\mu + B r_b^{4}}{ r_b^{2}} \,.
\]
Now, from equation (\ref{seed1}), the density profile is
\begin{equation}
\rho= \frac{3\mu- \left(5Br^2-3Br_b^2\right)r_b^2}{8\pi r_b^{2}} \,,
\label{deneos1}
 \end{equation}
and it is easy to see that:
 \begin{equation}
 \rho_c=\frac {3\left(\mu +Br_b^{4}\right)}{ 8 \pi r_b^{2}}\quad \mbox{and}\quad
 \rho_b= \dfrac{5\mu}{8\pi r_b^2} - \dfrac{2}{3}\rho_c \,,
 \label{rhobT7}
 \end{equation}
with the mass function expressed as
 \begin{equation}
m=  \frac{\mu}{2} \frac {{r}^{5}} {r_b^{4} } +\frac{4\pi\rho_c}{3}
 \left( 1-\frac {{r}^{2}}
 {r_b^{2}} \right) r^3 \,.
\label{MassSeed1} 
 \end{equation}
 Finally, the constant $B$ (equation (\ref{rhobT7})) is related to the quantity $\varkappa$, by
 \begin{equation}
  B= \frac{3\mu}{r_b^4}\frac{1-\varkappa}{3\varkappa+2} \,. 
 \end{equation}
\begin{figure}[t!]
\centering
\includegraphics[width=3.2in,height=2.4in]{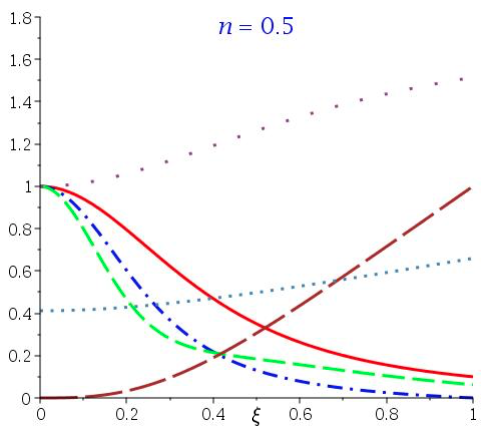}
\includegraphics[width=3.2in,height=2.4in]{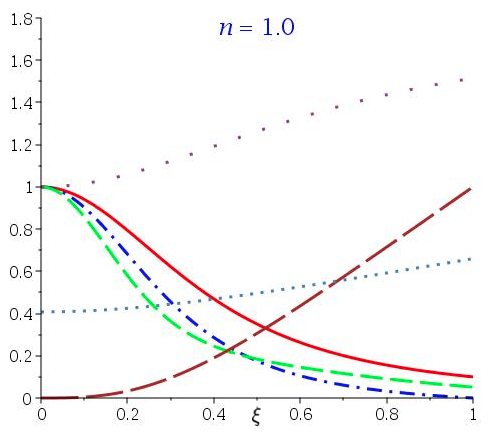}\\
\includegraphics[width=3.2in,height=2.4in]{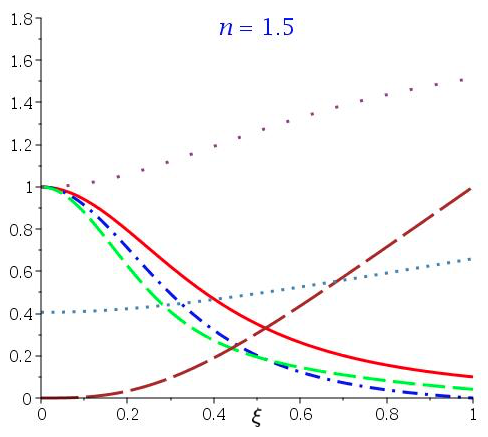}
\includegraphics[width=3.2in,height=2.4in]{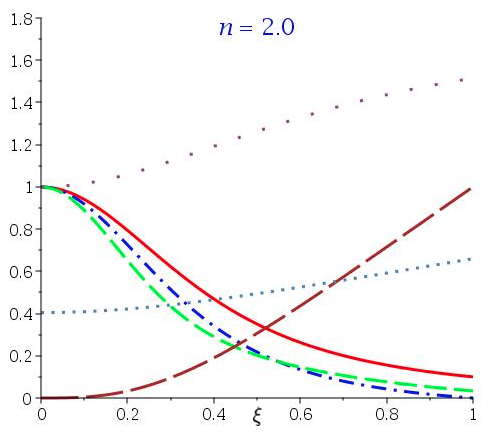}\\
\includegraphics[width=2.6in,height=0.4in]{Figures/legend2}
\caption{Physical variables and metric functions for a matter configuration with  EoS-2.  The different plates illustrate various values of $n$ having $\sigma= 0.12$. The normalised distributions of density, radial/tangential pressures, mass and the components of the metric tensor are given in terms of the dimensionless radius $\xi=r/r_b$. All the physical and geometrical variables are well behaved and comply with acceptability criteria {\bf C1}, {\bf C2},  and {\bf C3}. }
\label{EoS2Functions}
\end{figure}

Figure \ref{EoS1Functions} displays the profiles of the metric function and the physical variables for this EoS. The various plates illustrate different values of the polytropic index for a fixed stiffness parameter $\sigma~=~0.12$.

\subsection{Modeling the anisotropic, polytropic EoS-2}

The second $\lambda$-seed-function (\ref{seed2}) drives to an integral (\ref{nuMaster}) of the form
\begin{eqnarray}
\nu&=& -\left[
\frac {1+3\alpha}{4} + \frac{ A\left(1+\alpha\right)}{2B} -\frac{8\pi \beta}{B}\left[\frac12 - \frac{A}{B}\right] \right] \ln  \left[2+ B{r}^{2}\right] 
 -\frac{4\pi A\beta\,{r}^{2}}{B}+ \frac{\alpha \ln \left[1+A{r}^{2} \right]}{2}  \nonumber \\
& +& 4\pi \kappa \int r {\rm e}^{2\,\lambda}\, \rho ^{1+\frac{1}{n}} \,{\rm d}r +\mathcal{\mathbf{C}} \,.
\label{nueos2}
\end{eqnarray}
Again, the integration approach is in the appendix for $n = 0.5, 1.0, 1.5$ and $2.0$, and the conditions $P(r_b)=0$ and $m(r_b)=M$ lead to 
\[
\beta=k\rho_b^{1+\frac{1}{n}}+\alpha \rho_b \,, \quad A=\frac {2\mu +B r_b^2}{2 \left(1- \mu \right)r_b^2 } \,.
\]
\begin{figure}[t!]
\centering
\includegraphics[width=3.2in,height=2.6in]{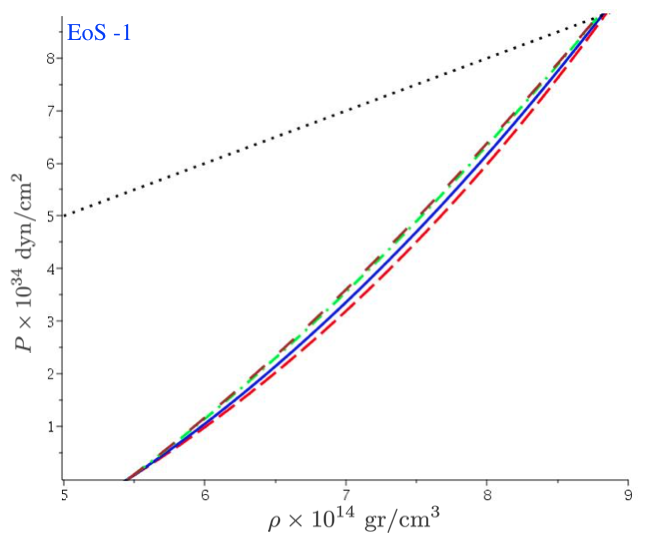} \hspace{0.3cm}
\includegraphics[width=3.2in,height=2.6in]{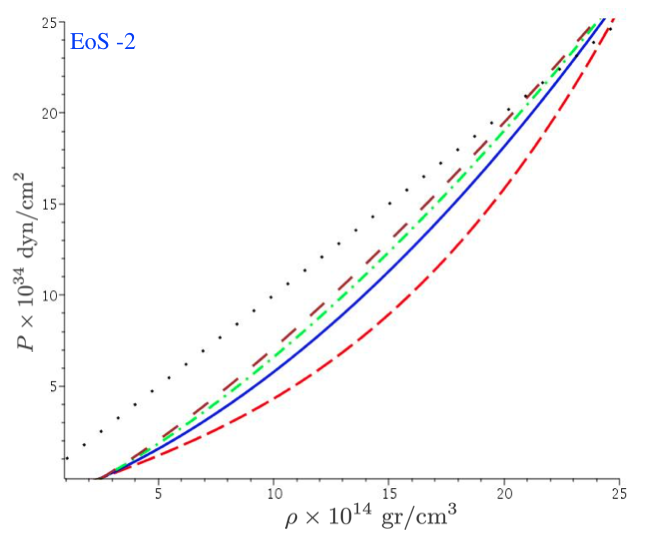}\\
\includegraphics[width=2.8in,height=0.28in]{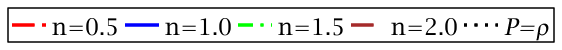}
\caption{The Radial pressure, $P$,  as a function of energy density, $\rho$, for different values of the polytropic index $n$ corresponding to an object of radius $r_b=10$~Km  and $\sigma=0.12$. The displayed profiles utilize the parameters listed in Table \ref{tab:PhysicalParameters}, and the dotted line represents the EoS stiff limit,  $P=\rho$. The resulting models for EoS-2 are stiffer than those emerging from EoS-1, and the higher the polytropic index is, the stiffer the model becomes.}
\label{FigPvsrho}
\end{figure}
Now, the density can be written as
\begin{equation}
\rho= \frac { \left[2+B r_b^2 \right]\left[ \left( 6[1-\mu]+B r^2 \right)r_b^2 +2\mu {r}^{2} \right] \mu} {8\pi \left[\left(2[1 -\mu]+B r^2\right) r_b^2 +2\mu {r}^{2} \right]^{2}}\,,
\label{deneos2}
 \end{equation}
while at the surface boundary we have
\begin{equation}
\rho_c= \dfrac{3\mu}{16\pi} \dfrac {2+B r_b^2 } 
 {(1 -\mu) r_b^2} \quad \mbox{and} \quad 
 \rho_b=\frac{\mu}{32\pi^2}\frac { 3\mu +4\pi r_b^{2}\rho_c}{r_b^{4}\, \rho_c} \,.
 \label{rhobBu}
\end{equation}
For this model, the mass is
\begin{equation}
m=\frac{4\pi\mu  \rho_c \, {r}^{3}}{  8 \pi  \rho_c r^2  +3 \mu\left(1 - \dfrac{r^2}{r_b^2} \right) } \,,
\label{MassSeed2}
\end{equation}
and the constant $B$ (equations (\ref{rhobBu}) )is given by 
\begin{equation}
B = \frac {(1-\mu) \sqrt { 24\varkappa+1}-\mu-6\varkappa +1 }{3\varkappa r_b^{2}} \,.
\end{equation}

Now, in Figure \ref{EoS2Functions} we plot the profiles for the metric function and the physical variables for this EoS2. Again, the various plates illustrate different polytropic indexes for a fixed stiffness parameter $\sigma~=~0.12$.

\section{Acceptability conditions for EoS-1 and EoS-2 models}
\label{AnalyticalAcceptability} 
In this section we shall show that, for various values of $n = 0.5, 1.0, 1.5$ and $2.0$,  there are some plausible EoS-1 and EoS-2 models, satisfying the acceptability conditions discussed in Section \ref{PhysicalAcceptabilityConditions}. The physical parameters are: $\mu$, $\varkappa$, $\alpha$ and $\sigma$, and Table \ref{tab:PhysicalParameters} displays the values of the chosen parameters to model physically significant anisotropic compact objects. 
\begin{table}[h!]
    \centering
    \begin{tabular}{|c|c|c|} \hline
       Input parameters  & EoS-1  & EoS-2 \\ \hline \hline
      $\mu= 2M/r_b$ & $0.43$ & $0.34$  \\ \hline
      $\varkappa= \rho_b/\rho_c$ & $0.60$ & $0.10$  \\ \hline
      $\alpha$ & $0.05$ ($n=0.5$) & $0.05$  \\  \hline 
    \end{tabular} 
     \begin{tabular}{|c|c|c|} \hline
         Output parameters & EoS-1  & EoS-2 \\ \hline \hline
      $\rho_c \times 10^{15}$ (gr/cm$^{3})$ & $0.91$ & $2.59$  \\ \hline
      $\rho_b \times 10^{14}$ (gr/cm$^{3})$ & $5.46$ & $2.59$  \\ \hline
      $M~(M_\odot)$ & $1.46$ & $1.15$  \\ \hline  
    \end{tabular}   
        \caption{Physical parameters for analytic polytropic solutions for EoS-1 and EoS-2, which fulfil the acceptability conditions. For EoS-1 the other $\alpha$ values correspond to:  $-0.10$ ($n=1.0$), $-0.15$ ($n=1.5$) and $-0.25$ ($n=2.0$). The parameter $\sigma$ can take values between $0.1$ and 0.18 approximately. The model EoS-1 could describe the mass of the millisecond pulsar in  SR J1738+0333. The mass for the EoS-2 compact object is close to the lowest-mass-pulsar J0453+1559 companion ($1.174 \pm 0.004~M_\odot$). }
    \label{tab:PhysicalParameters}
\end{table}

The anisotropic EoS-1 could resemble the mass millisecond pulsar in  SR J1738+0333. This binary system of a pulsar and a pulsating white dwarf has recently become a gravitational laboratory \cite{FreireEtal2012, KilicEtal2015, OzelFreire2016}. Now, regarding the EoS-2 model, it is also interesting to point out that those parameters could describe a low-mass pulsar.  The mass for the EoS-2 compact object ($1.15~M_\odot$) is close to the lowest-mass-pulsar J0453+1559 companion ($1.174 \pm 0.004~M_\odot$), the smallest precisely measured mass for any Neutron Star. 

Figure \ref{FigPvsrho} plots the  radial pressure $vs$ energy density, with densities corresponding to an object of radius $r_b=10$~Km  and $\sigma=0.12$ with $0.1 < \sigma < 0.18$, for both matter configuration. The models resulting from EoS-2 are stiffer than those from EoS-1.

The physical variables and the metric function distribution for the interior structures of ``master polytropic spheres'' are shown in Figures \ref{EoS1Functions} (EoS-1) and \ref{EoS2Functions} (EoS-2).  
 As stated in {\bf C1}, $g_{rr}~=~{\rm e}^{2\lambda} \geq 1$ having a maximum at the sphere boundary, while the $g_{tt}={\rm e}^{2\nu}$ component is always less than one and has a minimum at $\xi=0$. There are unstable polytropes in reference \cite{Tooper1964}, exhibiting a maximum for $g_{rr}$ at some point, $\xi$, within the sphere.  

In our case, for $\sigma = 0.12$, the density, the radial/tangential pressure distributions decrease rapidly as a function of the radius but always maintaining the conditions $\rho > P$ and $\rho > P_\perp$. This guarantees the fulfilment of {\bf C2}, {\bf C3}, and {\bf C4}.  For both equations of state, the tangential pressure at the boundary become larger as $n$ increases, which suggests that for higher $n$ values, i.e. $n> 2$, the strong energy condition, {\bf C4}, may not be satisfied.  

\begin{figure}[t!]
\centering
\includegraphics[width=3.2in,height=2.6in]{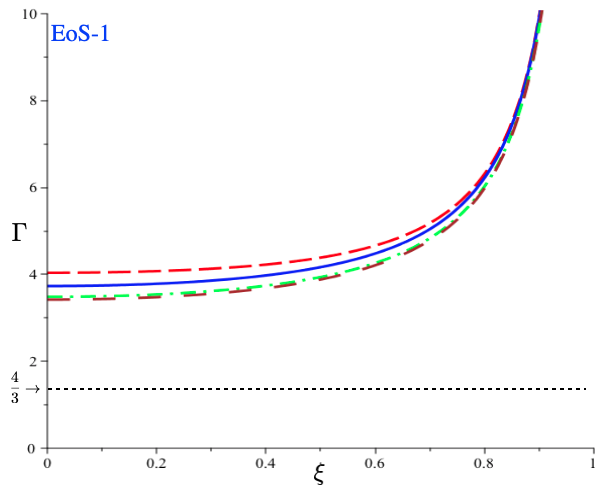} \hspace{0.3cm}
\includegraphics[width=3.2in,height=2.6in]{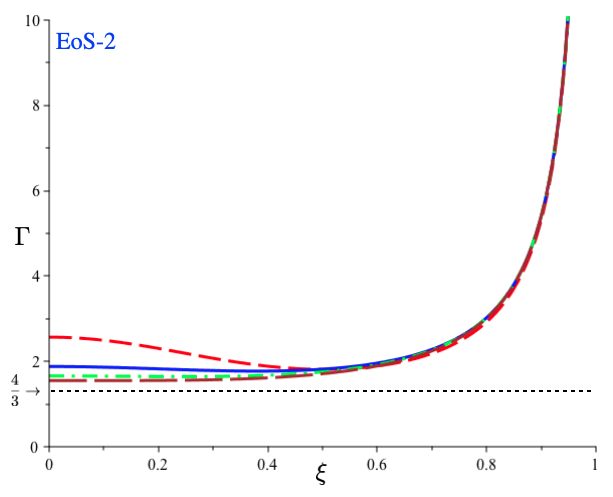}\\
\includegraphics[width=2.8in,height=0.25in]{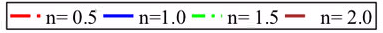}
\caption{The  adiabatic index $\Gamma$ as a function of the  dimensionless  radius $\xi=r/r_b$ for the EoS-1 (left) and EoS-2 (right), for different values of the index $n$ and $\sigma= 0.12$. The {\bf C5} condition is completely satisfied by both models.}   
\label{figC5}
\end{figure}

\begin{figure}[t!]
\centering
\includegraphics[width=3.2in,height=2.4in]{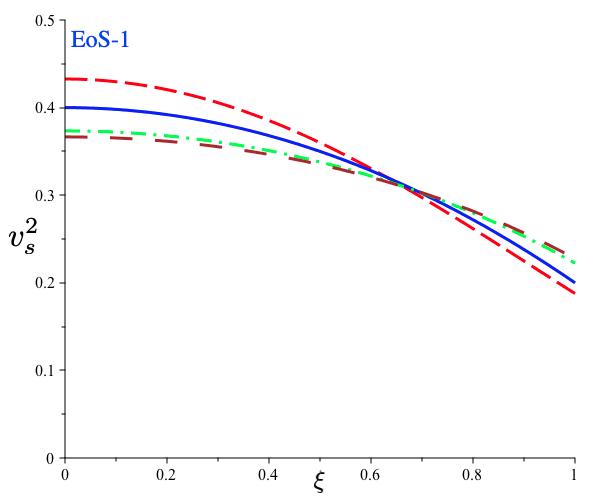} \hspace{0.3cm}
\includegraphics[width=3.2in,height=2.4in]{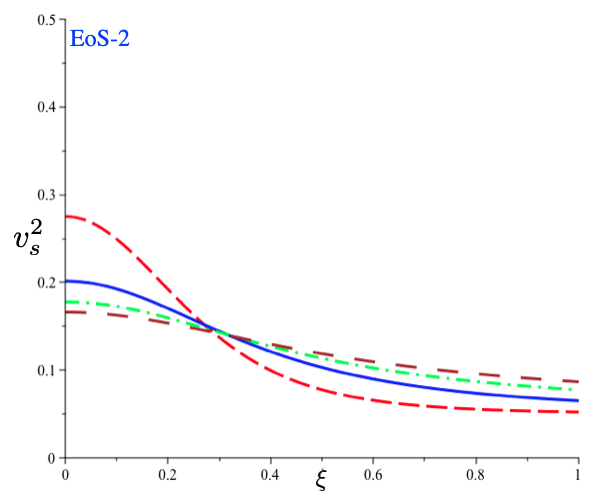}\\
\includegraphics[width=3.2in,height=2.4in]{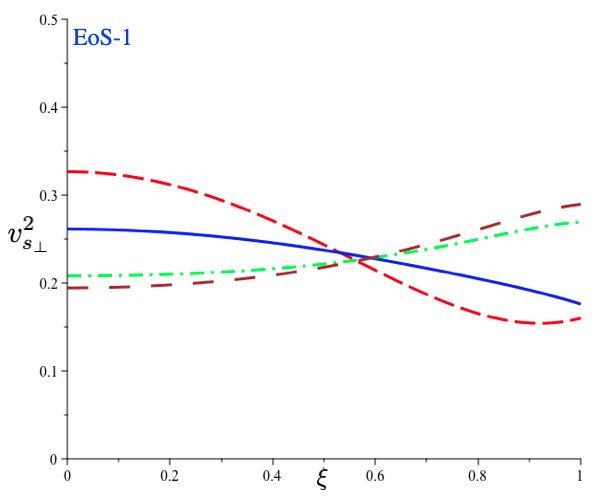} \hspace{0.3cm}
\includegraphics[width=3.2in,height=2.4in]{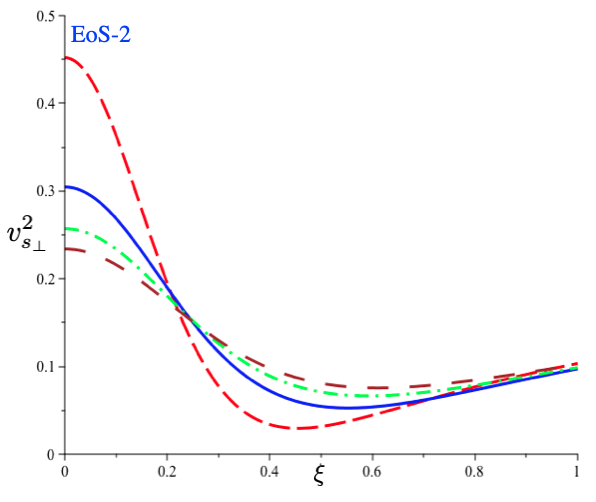}\\
\includegraphics[width=2.8in,height=0.25in]{Figures/legend3a}
\caption{Velocity of sound vs  the dimensionless radius $\xi=r/r_b$ 
for a matter configuration with master anisotropic polytropic: EoS-1 (left) and EoS-2 (right).  The upper plates show the radial speed of sound while the lower ones the tangential for different values of $n$ having $\sigma= 0.12$. In both cases, the radial sound velocity is a monotonically decreasing function.  Condition {\bf C6} is satisfied by not exceeding the velocity of light and the tangential velocity of sound  has a finite value on the surface of the star.}
\label{figC6}
\end{figure}

As Figure \ref{figC5} shows, the condition for the adiabatic index ({\bf C5}) is satisfied for both equations of state; while in Figure \ref{figC6} we display  condition {\bf C6}. That is, the minimum and maximum levels so that both the radial and tangential sound velocity do not exceed the velocity of light.

\begin{figure}[t!]
\centering
\includegraphics[width=3.2in,height=2.6in]{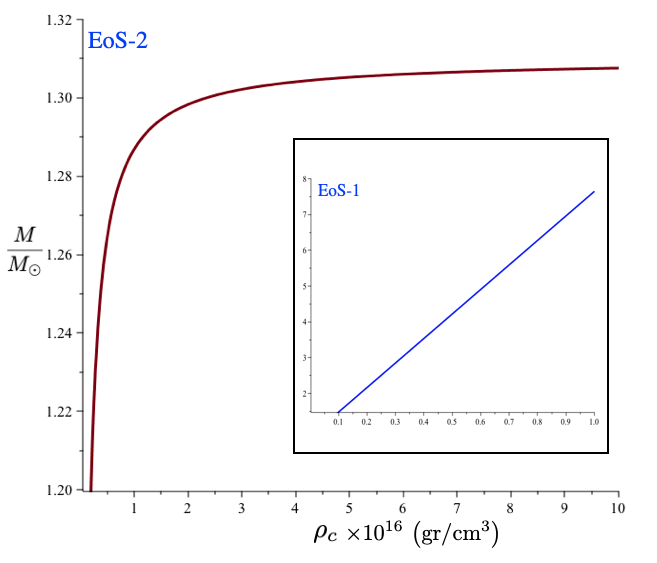} \hspace{0.3cm}
\includegraphics[width=3.2in,height=2.6in]{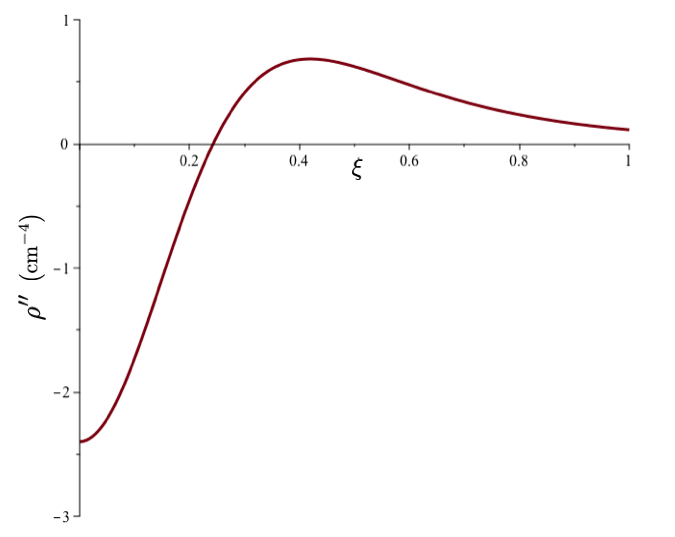}
\caption{The plate on the left shows the total mass function {\it vs} central density for the EoS-2 model and in the inset the EoS-1 case. The plate on the right shows the $\rho''$ {\it vs} the dimensionless radius $\xi=r/r_b$ for the EoS-2 model, on the other hand, $\rho''=-0.054$ cm$^{-4}$ is constant for the EoS-1 model.}
\label{figC7C9}
\end{figure}

Conditions {\bf C7} and {\bf C9} have to do exclusively with the intrinsic properties of seed functions. In the case of {\bf C7}  we can observe from equations (\ref{MassSeed1}) and (\ref{MassSeed2}) that they are the corresponding mass functions for the equations of state EoS-1 and EoS-2. By inspection, from equation (\ref{MassSeed1}) it is easy to see that $M$ is a linear function of the central density $\rho_c$, while equation (\ref{MassSeed2}) has an asymptotic behaviour towards a value of $\approx 1.31$ $M_\odot$ (see the left plate of Figure \ref{figC7C9}). 

On the other hand, for condition {\bf C9} ($\rho''<0$) it is easy to appreciate  that for EoS-1:
\[
\rho''(r)= -\frac{5B}{4\pi}= \mbox{constant} <0\,,
\]
since $B>0$. With the values of the parameters in Table \ref{tab:PhysicalParameters},  $\rho''=-0.054$ cm$^{-4}$, while for EoS-2 we have:
\[
\rho''(r)= \frac{ A\left(2A-B\right)\left(3A^2r^4+22Ar^2-5 \right)   }{8\pi(1+Ar^2)^4} \,,
\]
whose profile is ploted on the right side of Figure \ref{figC7C9}. Clearly, EoS-1 fulfils the adiabatic convection criterion {\bf C9} but EoS-2 is only stable in a region near the nucleus.  

\begin{figure}[b!]
\centering
\includegraphics[width=3.2in,height=2.6in]{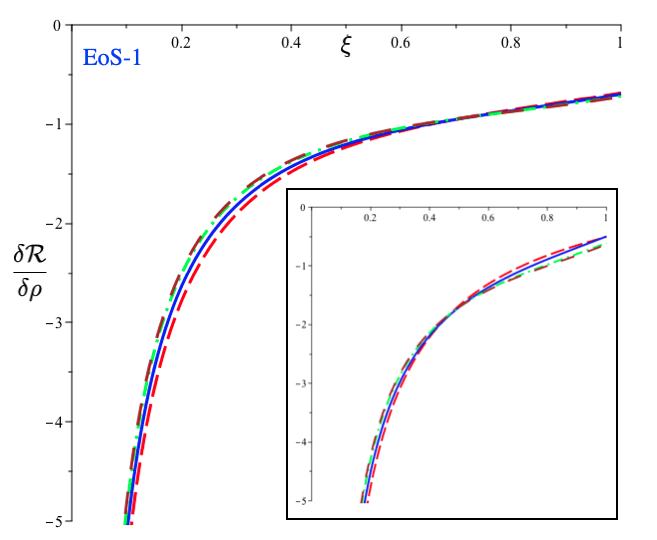} \hspace{0.3cm}
\includegraphics[width=3.2in,height=2.6in]{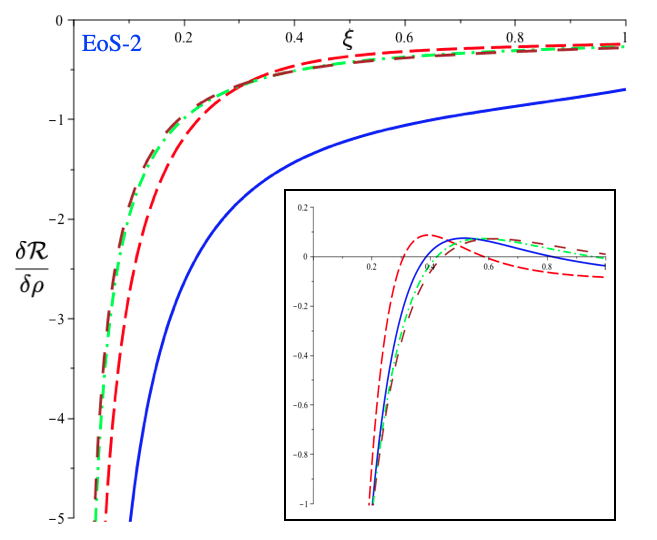}\\
\includegraphics[width=2.8in,height=0.25in]{Figures/legend3a}
\caption{The function $\delta \mathcal{R}/\delta \rho$  {\it vs}  the dimensionless radius $\xi=r/r_b$  for a matter configuration with master anisotropic polytropic: EoS-1 (left) and EoS-2 (right) for different values of $n$ and $\sigma= 0.12$. The same function is shown in the small insets but with the pressure gradient zero, i.e. $\delta  \mathcal{R}_p = 0$.}
\label{figC8}
\end{figure}

Concerning the cracking instability, condition {\bf C8}, we plot $\delta \mathcal{R}/\delta \rho$ in Figure \ref{figC8}. The small plate displays the same function with a vanishing pressure gradient, i.e,  $\delta   \mathcal{R}_p = 0$. It is clear that when density perturbations do not affect the pressure gradient, the $\delta \mathcal{R}$-sign can change and potential cracking instabilities may appear. However, if  the gradient reacts to the perturbation, $\mathcal{R}_p\neq0$ then $\mathcal{R}$ does not change sign and the matter configuration becomes stable against cracking \cite{GonzalezNavarroNunez2017}. In our modeling EoS-1 is stable to local density perturbations, while EoS-2 is not.

\section{Conclusions and final remarks}
\label{FinalRemarks}
We checked most of the models for polytropic anisotropic relativistic spheres encountered in the literature and found several misunderstanding. We have noticed that some anisotropic polytropic models may have singular tangential sound velocity for polytropic indexes greater than one and this is overlooked in several papers (see, for example, references \cite{ThirukkaneshRagel2012, Ngubelanga2015, AbellanEtal2020, TakisaMaharaj2013, Malaver2015, NgubelangaMaharaj2017, SharifSadiq2018}). It is a general result when employing the ``standard'' polytropic EoS together with an ansatz on the metric functions.  This pathology is not present in polytropes when other strategies are implemented obtaining the anisotropic pressure~\cite{DonevaYazadjiev2012, HerreraBarreto2013, RaposoEtal2019, AbellanEtal2020}.

We implemented a generalization to the polytropic equation of state in terms of physically meaningful parameters:
\begin{itemize}
    \item $\sigma = P_c/\rho_c$,  describing the stiffness at the centre of the matter distribution;
    \item $\varkappa =  \rho_b/\rho_c$, sketching the density drop from the centre to the surface of the compact object,
    \item and $\alpha$, related to the causality of the radial sound velocity.
\end{itemize}

We found two new analytical anisotropic solutions for the ``master'' EoS starting from intuitive ansatz for the density profiles, avoiding cumbersome and redundant auxiliary variables. Both solutions were obtained by taking one of the metric functions as a seed function to integrate equation (\ref{Integnu})  via a barotropic equation of state, i.e. equation (\ref{MasterPolytropic}). We evaluated the relativistic master polytropic equation for different values of index $n$: $0.5$, $1.0$, $1.5$ and $2.0$. 

We sketched an algorithm to generate exact anisotropic solutions starting from a barotropic EoS and by choosing a particular guess on the form of one of the metric functions to close the system of Einstein's equations. Any barotropic EoS, together with a density profile, could feed equation \ref{Integnu} to obtain the other metric function. In our scheme, it is not necessary to introduce any changes of variables to expedite a possible analytical integration, and most of these substitutions (also employed in references \cite{FerozeSiddiqui2011, Malaver2015, ThirukkaneshMaharaj2008, SinghEtal2020}) appears to be redundant. The models discussed this work do not lack physical meaning, and we list several candidates that can be described within this anisotropic-polytropic framework. We have also checked the acceptability conditions for these two new solutions and found that the EoS-1 is stable for the whole set of nine criteria.

\section*{Acknowledgments}
We gratefully acknowledge the financial support of the Vicerrector\'ia de Investigaci\'on y Extensi\'on de la Universidad Industrial de  Santander and the financial support provided by COLCIENCIAS under project No. 8863.

\section*{Appendix}
\label{Appendix1}
\subsection*{A-1. Polytropic integrals}
In equations (\ref{nueos1}) and (\ref{nueos2}) the integral 
\begin{equation}
 I_p=4\pi \kappa \int r{{\rm e}^{2\,\lambda  }} \rho ^{1+\frac{1}{n}}\,{\rm d}r +\mathcal{\mathbf{C}} \,,
\end{equation}
it was indicated. Here we show the solutions for different values of $n$.

{\bf EoS-1:}

 $\bullet \,\, n=\frac12$:
  \begin{equation*}
   I_p= -\frac{\kappa}{512 \pi ^2}   \left[ 25 B{r}^{2} \left[ 5\,B{r}^{2}+8\,A \right]   +  5  \left[ 7\,{A}^{2}-25\,B \right] {\rm e}^{-2\lambda}  - \frac {2A \left[ 19\,{A}^{2}-75\,B \right] }{\sqrt {{A}^{2}-4\,B}}\mathcal{Y} \right]\,.
\end{equation*}
 
 $\bullet \,\, n=1$
  \begin{equation*}
  I_p=\frac{\kappa}{64\pi} \left[50B{r}^{2} + 5A {\rm e}^{-2\lambda}  -\frac { 2\left[ 13\,{A}^{2}-50\,B
 \right] }{\sqrt {{A}^{2}-4\,B}} \mathcal{Y}\right] \,,
 \end{equation*}
where
 \begin{equation*}
\mathcal{Y}= {\rm arctanh}\left[\frac{A+2 B r^2}{\sqrt{A^2-4
   B}}\right] \,.
 \end{equation*}

$\bullet \,\,  n=\frac32$

For this case and the next case we will define the following two auxiliary variables
\begin{equation*}
  \mathcal{X}_1 =  5\sqrt{A^2-4 B} \quad \mbox{and} \quad  \mathcal{X}_2 =  {-5 B r^2-3 A} \,,
\end{equation*}
So
\begin{equation*}
  I_p=  \frac{ 5\kappa}{128 \pi^{\frac23} \mathcal{X}_1} \left[ 2 \sqrt[3]{2} \sqrt{3} \left( \left[-A- \mathcal{X}_1\right]^{\frac53} \left[\mathcal{T}_1 -\mathcal{T}_2\right]  \right) -\sqrt[3]{2}\  \mathcal{T} - 12 \mathcal{X}_1
  \mathcal{X}_2^{\frac23}\right] \,,
\end{equation*}
where
\begin{eqnarray*}
\mathcal{T}_1 =   {\rm arctan}\left[\frac{2 \sqrt[3]{2\mathcal{X}_2}  }{\sqrt{3} \sqrt[3]{-A- \mathcal{X}_1 }}  + \frac{1}{\sqrt{3}}\right]  \,, \quad
\mathcal{T}_2 = {\rm arctan}\left[\frac{2 \sqrt[3]{2 \mathcal{X}_2}  }{\sqrt{3} \sqrt[3]{-A+ \mathcal{X}_1 }}  + \frac{1}{\sqrt{3}}\right] 
\end{eqnarray*}
and
\begin{eqnarray*}
\mathcal{T} &=&  -2 \ln
   \left[\frac{\sqrt[3]{-A-\mathcal{X}_1 }}{\sqrt[3]{2}}-\sqrt[3]{\mathcal{X}_2}\right]
   \left[-A- \mathcal{X}_1\right]^{\frac53} + 2 \left[-A+\mathcal{X}_1\right]^{\frac53}
   \ln \left[\frac{\sqrt[3]{-A+\mathcal{X}_1 }}{\sqrt[3]{2}}- \sqrt[3]{\mathcal{X}_2} \right]\\
   &+& \left(-A-\mathcal{X}_1 \right)^{\frac53} \ln
   \left[\frac{\left(-A-\mathcal{X}_1\right)^{\frac23}}{\sqrt[3]{4} }+\frac{\sqrt[3]{\mathcal{X}_2}
   \sqrt[3]{-A-\mathcal{X}_1 }}{\sqrt[3]{2}}+\mathcal{X}_2^{\frac23}\right]  \\
   &-& \left(-A+\mathcal{X}_1\right)^{\frac53} \ln
   \left[\frac{\left(-A+\mathcal{X}_1 \right)^{\frac23}}{\sqrt[3]{4}}+\frac{ \sqrt[3]{\mathcal{X}_2} \sqrt[3]{-A+\mathcal{X}_1 }}{\sqrt[3]{2} }+\mathcal{X}_2^{\frac23}\right] \,.
\end{eqnarray*}

$\bullet \,\,  n=2$
\begin{eqnarray}
  I_p=- \frac{5 \kappa}{ 4 \sqrt{2\pi}}\left[
 \frac{\left[ A\mathcal{X}_1 +13 A^2 - 50 B\right] \mathcal{Y}_1 }{\sqrt{2} \mathcal{X}_1 \sqrt{-A -\mathcal{X}_1 }} + \frac{\left[ A \mathcal{X}_1 - 13 A^2 + 50
   B\right] \mathcal{Y}_2 }{\sqrt{2} \mathcal{X}_1 \sqrt{-A+\mathcal{X}_1}} + \sqrt{\mathcal{X}_2 }\right]\,,
\end{eqnarray}
where  
   \begin{equation*}
\mathcal{Y}_1=  {\rm arctanh}  \left[\frac{\sqrt{2 \mathcal{X}_2}
    }{\sqrt{-A -\mathcal{X}_1 }}\right]
\quad \mbox{and} \quad 
\mathcal{Y}_2= {\rm arctanh}  \left[\frac{\sqrt{2 \mathcal{X}_2}  }{\sqrt{-A+\mathcal{X}_1}}\right]\,.
  \end{equation*}

 {\bf EoS-2:}

$\bullet \,\, n=\frac12$:
 \begin{eqnarray*}
I_p &=&  -\frac{\kappa} {768 \pi ^2 K^2 } \left[ \frac{3 \left(3 A^2 K^2-12 A B K+13 B^2\right) }{\left(Ar^2+1\right)^2}+\frac{6 (B-AK)^2}{\left(Ar^2+1\right)^4} \right. \\ 
 &+& \left. \frac{4 (5 B-3 A K) (B-AK)}{\left(A r^2+1\right)^3}+\frac{3 (3 B-A K)^3}{\left(A r^2+1\right)(B-A K)}\right. \\ 
 &-& \left. \frac{3 B (3 B-A K)^3}{(B-A K)^2} \ln \left[\frac{A r^2+1}{B r^2+K}\right] 
 \right] \,.
 \end{eqnarray*}

 $\bullet \,\, n=1$:
 \begin{equation*}
I_p= \frac{\kappa}{32 \pi K} \left[\frac{2 \left[ B
   \left(4 A r^2+5\right)-A K \left(2 A
   r^2+3\right)\right] }{\left(A r^2+1\right)^2} + \frac{(A K-3B)^2}{AK - B} \ln \left[\frac{A r^2+1}{B
   r^2+K}\right ]\right]\,.
 \end{equation*}

 $\bullet \,\, n=\frac32$:
\begin{eqnarray*}
I_p &=& -\frac{3\kappa\left[\frac{\left(A r^2+3\right) (A
   K-B)}{K \left(A r^2+1\right)^2}\right]^{\frac23}
   }{160 \pi ^{2/3} B \left(A
   r^2+3\right)} \left[ 
   \left[{\frac{2}{A r^2+1}+1}\right]^{\frac13} \left[2 (3 A K-7 B)
  \mathcal{F}   - 5 \left(A r^2+1\right) (2 B-A K)
  \mathcal{G} \right] \right.\\
  &+& \left. 5 B \left(A r^2+3\right)
   \right] \,,
\end{eqnarray*}
where 
\begin{eqnarray*}
\mathcal{F} &=&  
   F_1\left(\frac{5}{3} ; \frac{1}{3},1;\frac{8}{3};-\frac{2
   }{A r^2+1},\frac{B-A K}{A B r^2+B}\right) \,, \\
\mathcal{G} &=&  F_1\left(\frac{2}{3};\frac{1}{3},1;\frac{5}{3};-\frac{2}{A r^2+1},\frac{B-A K}{A B r^2+B}\right) \,.
\end{eqnarray*}
Here $F_1$ is the Appell hypergeometric function of two variables  $F_1 (a; b_1,b_2;c;x,y)$.

 $\bullet \,\, n=2$:
\begin{equation*}
 I_p= \frac{ \kappa  \sqrt{{\left(A r^2+3\right) (AK-B)}} \left[\left(A  r^2+1\right) \mathcal{T}  -2 \sqrt{B}  \sqrt{A r^2+3} \sqrt{3 B-A K} (AK-B) \right] } {8 \sqrt{KB} (A r^2+1) \sqrt{A r^2+3} (AK-B) \sqrt{6 \pi  B-2 \pi  A K}}\,,
\end{equation*}
 where
\begin{eqnarray*}
\mathcal{T} &=& \sqrt{2 B} (7 B-3 A K)
   \sqrt{3 B-A K} {\rm arctanh}\left(\frac{\sqrt{A
   r^2+3}}{\sqrt{2}}\right) \\
   &-& 2 (A K-3 B)^2 {\rm arctanh}\left(\frac{\sqrt{B} \sqrt{A r^2+3}}{\sqrt{3B-A K}}\right)\,.
\end{eqnarray*}

\subsection*{A-2. Cracking against local perturbation}

Just for completeness we shall consider in this Appendix the local perturbations of density, $\delta \rho~=~\delta \rho(r)$, and show the difference between the present  {\bf C8} and the previous more simple cracking criterion~\cite{AbreuHernandezNunez2007b}. The $\delta \rho(r)$ fluctuations induce variations in all the other physical variables, i.e. $m(r), P(r), P_\perp(r)$ and their derivatives, generating a non-vanishing total radial force distribution. For further details, we refer interested readers to~\cite{GonzalezNavarroNunez2017,GonzalezNavarroNunez2015,HernandezNunezVasquez2018} and references therein. 

Following \cite{GonzalezNavarroNunez2017}, we formally expand the TOV equation (\ref{TOVStructure1}) as:
\begin{equation}
 \label{Ranitov}
 \mathcal{R} \equiv \frac{\mathrm{d} P}{\mathrm{d} r} +(\rho +P)\frac{m + 4 \pi r^{3}P}{r(r-2m)} -\frac{2}{r}\left(P_\perp-P \right) \, ,
\end{equation}
as
\begin{equation}
\label{RAniExpanded}
\mathcal{R} \approx \mathcal{R}_{0}(\rho, P, P_\perp, m, P^{\prime}) +
\frac{\partial \mathcal{R}}{ \partial \rho} \delta \rho
+\frac{\partial \mathcal{R}}{ \partial P} \delta P
+ \frac{\partial \mathcal{R}}{\partial P_\perp} \delta P_\perp
+\frac{\partial \mathcal{R}}{ \partial m} \delta m 
+\frac{\partial \mathcal{R}}{ \partial P^{\prime}} \delta P' \,,
\end{equation}
where $\mathcal{R}_{0}(\rho, P, P_\perp, m, P^{\prime})=0$, because initially the configuration is in equilibrium.

Accordingly, local density perturbations, $\rho  \rightarrow \rho + \delta \rho$, generate fluctuations in mass, radial pressure, tangential pressure and radial pressure gradient, that can be represented up to linear terms in density fluctuation as: 
\begin{eqnarray} 
 \delta P &=& \frac{\mathrm{d}P}{\mathrm{d} \rho} \delta \rho =  v_s^2 \delta \rho \,, 
 \label{deltas1} \\   \nonumber \\
 \delta P_\perp &=& \frac{\mathrm{d}P_\perp}{\mathrm{d} \rho} \delta \rho =  v_{s_\perp}^2 \delta \rho \,, 
 \label{deltas2} \\   \nonumber \\ 
\delta P' &=& \frac{\mathrm{d} P'}{\mathrm{d} \rho} \delta \rho  =   \frac{\mathrm{d \ }}{\mathrm{d} \rho} \left[ \frac{\mathrm{d} P}{\mathrm{d}r} \right] \delta \rho =   \frac{\mathrm{d}}{\mathrm{d}\rho} \left[ \frac{\mathrm{d}P}{\mathrm{d}\rho} \frac{\mathrm{d}\rho}{\mathrm{d} r} \right]\delta \rho =
 \frac{\mathrm{d}}{\mathrm{d}\rho} \left[v_s^2 \rho' \right]\delta \rho =
 \frac{1}{ \rho'} \frac{\mathrm{d}}{\mathrm{d}r} \left[v_s^2 \rho' \right]\delta \rho 
  \label{deltas3} \nonumber 
 \\   \nonumber \\
&=&   \left[  (v_s^2)' + v_s^2\frac{\rho''}{\rho'} \right] \delta \rho  \,,  \\  \nonumber \\
\delta m &=& \frac{\mathrm{d} m}{\mathrm{d} \rho} \delta \rho =  \frac{\mathrm{d} m}{\mathrm{d} r} \left( \frac{\mathrm{d} r}{\mathrm{d} \rho} \right) \delta \rho   =  \frac{m'}{ \rho'} \delta \rho = \frac{4 \pi r^2 \rho}{\rho'}\delta \rho   \,.
\label{deltas4} 
\end{eqnarray}
where 
\begin{equation}
 v_s^{2} = \frac{\mathrm{d} P }{ \mathrm{d} \rho}  \qquad \mathrm{and} \qquad 
 v_{s_\perp}^2 = \frac{\mathrm{d} P_\perp }{ \mathrm{d} \rho}\,,
\end{equation}
are the radial and tangential sound speeds, respectively. 

Next, by using (\ref{deltas1})-(\ref{deltas4}) the above equation (\ref{RAniExpanded}) can be reshaped as:
\begin{equation}
 \delta \mathcal{R} \equiv \delta \underbrace{P'}_{\mathcal{R}_p}  + \delta \underbrace{\left[ (\rho +P)\frac{m + 4 \pi r^{3}P}{r(r-2m)} \right]}_{\mathcal{R}_g}  + \delta \underbrace{\left( 2\frac{P}{r}- 2\frac{P_\perp}{r}\right)}_{\mathcal{R}_a} = \delta \mathcal{R}_p + \delta \mathcal{R}_g +\delta \mathcal{R}_a \, ,
 \label{PartdeltaR}
\end{equation}
where it is clear that the density perturbations $\delta \rho(r)$ are influence the distribution of reacting pressure forces $\mathcal{R}_p$, gravity forces $\mathcal{R}_g$ and anisotropy forces $\mathcal{R}_a$.  Depending on this effect, each perturbed distribution force can contribute in a different way to the change of sign of $\delta \mathcal{R}$: each term can be written as 
\begin{equation}
\label{Fp_pg}
\delta \mathcal{R}_p = \left( \frac{P''}{\rho'} \right) \delta \rho = \left((v_s^2)' + v_s^2\frac{\rho''}{\rho'}\right) \delta \rho \, , \quad
\delta \mathcal{R}_g = \left( \frac{ \partial \mathcal{R}_g }{ \partial \rho } + \frac{ \partial \mathcal{R}_g }{ \partial P } v_s^2 + \frac{ \partial \mathcal{R}_g }{ \partial m } \frac{4 \pi r^2 \rho}{\rho'}\right) \delta \rho \quad \textrm{and}
\end{equation}
\begin{equation}
\label{Fa}
\delta \mathcal{R}_a = \left( \frac{ v_s^2 - v_{s_\perp}^2 }{ r } \right) \delta \rho ,
\end{equation}
with
\begin{equation}
\label{Fg_rho}
\frac{\partial \mathcal{R}_g}{\partial \rho}  =  \frac{m+  4 \pi r^3 P }{ r(r - 2m)}, \quad
\frac{\partial \mathcal{R}_g}{\partial P} = \left[ \frac{ m + 4 \pi r^3 ( \rho + 2 P )}{r(r-2m)} \right]
\; \; \textrm{and} \; \;
\frac{\partial \mathcal{R}_g}{\partial m}  =  \left[\frac{ (\rho + P)( 1 + 8\pi r^2 P)  }{(  2 m  - r )^2 }\right].
\end{equation}
Notice that if, as in \cite{AbreuHernandezNunez2007b}, the perturbation $\delta \rho$ is constant and does not affect the pressure gradient, we have: $\delta \mathcal{R}_p = 0$, 
\begin{equation}
\label{AbreuEtAl}
\delta \tilde{\mathcal{R}_g} = \left(2\frac{ m + 4 \pi r^3 ( \rho + 2 P )}{r(r-2m)} +\frac{4 \pi r^2}{3} \frac{ (\rho + P)( 1 + 8\pi r^2 P)  }{(  2 m  - r )^2}\right)\delta \rho, \quad \textrm{and } \;
\delta \tilde{\mathcal{R}_a} = \left( \frac{ v_s^2 - v^2_{s_\perp}}{ r } \right) \,.
\end{equation}
Thus, only anisotropic matter distribution can present cracking instabilities because $\delta \tilde{\mathcal{R}_g} >0$ for all $r$ and the possible change of sign for $ \delta \mathcal{R}$ should emerge from $\delta \mathcal{R}_a$ and the criterion against cracking is written as: 
\begin{equation}
\label{CrackingCriterion}
-1 \leq v_{s_\perp}^2 -v_s^2 \leq 0 \quad \Leftrightarrow \quad 
0 \geq \frac{\mathrm{d} P_\perp}{\mathrm{d} r} \geq \frac{\mathrm{d} P}{\mathrm{d} r} \,.
\end{equation}

\bibliographystyle{unsrt}
\bibliography{BiblioLN1909.bib}

\end{document}